\documentclass[aps,prl,twocolumn,superscriptaddress,showpacs,english]{revtex4-1}

\usepackage[ruled,vlined]{algorithm2e}
\usepackage{algorithmic}

\usepackage{float}
\usepackage{caption}
\usepackage{subcaption}
\usepackage[T1]{fontenc}
\usepackage[latin9]{inputenc}
\usepackage{babel}
\usepackage{amsmath}
\usepackage{amssymb}
\usepackage{wasysym}
\usepackage{graphicx}
\usepackage{multirow}
\usepackage{gensymb}
\usepackage[dvipsnames]{xcolor}
\usepackage{pifont}
\usepackage{microtype}
\usepackage{xcolor}
\usepackage{soul}
\usepackage[version=3]{mhchem} 
\usepackage{chemformula}[2014/04/08] 
\usepackage{booktabs}    
\usepackage{dcolumn}     
\usepackage{bm}          
\usepackage[linktocpage=true,
  colorlinks=true, 
  pdfborder={0 0 0},
  linkcolor=blue,
  citecolor=red,
  filecolor=yellow,
  urlcolor=blue,
  bookmarks,
  pdfauthor={},
]{hyperref}

\begin{document}

\title{Manifolds of quasi-constant SOAP and ACSF fingerprints and the resulting failure 
to machine learn four-body interactions}  

\author{Behnam Parsaeifard}
\affiliation{Department\ of\ Physics,\ University\ of\ Basel,\ Klingelbergstrasse\ 82,\ CH-4056\ Basel,\ Switzerland}

\author{Stefan Goedecker}
\affiliation{Department\ of\ Physics,\ University\ of\ Basel,\ Klingelbergstrasse\ 82,\ CH-4056\ Basel,\ Switzerland}



\begin{abstract}
Atomic fingerprints are commonly used for the characterization of local environments of atoms in machine learning and other contexts. In this work, we study the behaviour of two widely used fingerprints,  
namely the smooth overlap of atomic positions (SOAP) and the atom-centered symmetry functions (ACSF), 
under finite changes of atomic positions and demonstrate the existence of manifolds of quasi-constant fingerprints. These manifolds are found numerically by following eigenvectors of the sensitivity matrix  with quasi-zero eigenvalues. The existence of such manifolds in ACSF and SOAP causes a failure to machine learn four-body interactions such as torsional energies that are part of standard force fields. No such manifolds can be found for the Overlap Matrix (OM) fingerprint due to its intrinsic many-body character. 
\end{abstract} 

\maketitle

\section{introduction}
Atomic environment fingerprints encode information about the chemical environment such as bond-lengths to neighboring atoms or coordination numbers \cite{mallat, christensen2020fchl, smith2020ani,doi:10.1063/5.0015571, P3136,P5292,P5075,P5420,P4862,P4644,P5645,doi:10.1021,huang2017dna,mallat,huan}.
The Cartesian coordinates of the atoms in a system are not a useful fingerprint since 
the invariance of the energy under certain operations is not encoded in such a fingerprint.
 A fingerprint should be invariant under uniform translations, rotations, and permutation of identical atoms in the system.
 In addition fingerprints should be unique in the sense that two environments are
 guaranteed to be identical if their fingerprints are identical. If this condition is not satisfied two different non-degenerate structures will 
 be assigned the same energy by a machine learning scheme where such a fingerprint
 is used as an input. In the numerical context an even stronger condition has to be fulfilled. To facilitate machine learning schemes, the fingerprint difference has to correlate with the amount of dissimilarity
 between the environments. In particular this means that two structures 
 whose fingerprint difference is very small should also be very similar.
 We will show in this investigation that this condition is not always fulfilled for some standard fingerprints. 
 There is obviously no unique measure of
 dissimilarity, but the quality of the fingerprint can indirectly be deduced from the fact that different fingerprints lead
 to learning curves that have different convergence rates~\cite{langer2021representations,poelking2021benchml}. 
 As we will show the extremely small variation of some standard fingerprints on the manifolds we discovered  leads to a failure to machine learn certain four-body interactions.

 Two well-known fingerprints that are commonly used in machine learning of the potential energy surface are the smooth overlap of atomic positions (SOAP) \cite{bartok2013representing} and the atom-centered symmetry functions (ACSF). ACSFs consist of radial and angular symmetry functions. The Radial symmetry functions ($G_2$) are sums of two-body terms and describe the radial environment of an atom. The angular symmetry functions contain the summation of three-body terms and describe the angular environment of an atom~\cite{behler2007generalized,behler2011atom}.
 

  In the SOAP fingerprint,
  a Gaussian is centered on each atom within the cutoff distance around the reference atom $k$. The resulting density of atoms multiplied with a cutoff function, which goes smoothly to zero at the cutoff radius over some characteristic width,
  is then expanded in terms of orthogonal radial functions $g_{n}(r)$ and spherical harmonics $Y_{lm}(\theta,\phi)$ as $\rho^k(\mathbf{r})=\sum_{nlm}{c^k_{nlm}g_n(r)Y_{lm}(\theta,\phi)}$. The vector containing all $p^k_{nn'l}$'s, defined as $p^k_{nn'l}=\sqrt{\frac{8\pi^2}{2l+1}}\sum_{m}{c^k_{nlm}(c^{k}_{n'lm})^{*}}$, with $n,n' \leq n_{max}$ and $l \leq l_{max}$ is the SOAP fingerprint vector of atom $k$ ~\cite{bartok2013representing}. 
  
  The Overlap Matrix (OM) fingerprint is based on a diagonalization. 
  In the OM scheme we place a minimal basis set of Gaussian orbitals on all atoms within a cutoff radius and then calculate the overlap matrix between all the orbitals. The resulting overlap matrix is modulated by some smooth cutoff function. The eigenvalues of the resulting matrix form the fingerprint vector of the reference atom in OM ~\cite{zhu2016fingerprint, sadeghi2013metrics}. 
  { color{blue} In contrast to another method which is based on eigenvalues of a matrix, namely the Coulomb Matrix
method~\cite{coulomb}, our fingerprint consists of 4N eigenvalues and contains therefore enough information 
to specify the 3N Cartesian coordinates of an environment of N atoms~\cite{Moussa}.}
  We will show that both SOAP and ACSF 
  fingerprints are insensitive to certain movements, which allows us to construct manifolds of quasi-constant fingerprint for both.
  The QUIPPY~\cite{quippy} software, which is the original  software of the 
  SOAP developers, is used to calculate both the ACSF and SOAP fingerprints. 
  
  In a recent investigation~\cite{pozdnyakov2020incompleteness} of the SOAP fingerprint it was found that for the $CH_4$ molecule there are two distinct configurations which give rise to exactly the same fingerprint. In this work we go one step further and show that there exist even manifolds in configuration space with quasi-constant fingerprints.
  We call a fingerprint quasi-constant if its variation is so small that is behaves 
  numerically like a constant fingerprint.
  
\section{methodology}\label{methodology}

The sensitivity matrix was introduced to study the behaviour of atomic fingerprints under infinitesimal changes on the atomic coordinates \cite{parsaeifard2020assessment}. The square of fingerprint distance between a reference atomic configuration, $\bm R_0$, and a configuration displaced by $\bm{\Delta R}$, $\bm R$, can be expanded in a  Taylor series as: 

\begin{equation}
\left( \bm{F(R)} - \bm{F^0} \right)^2 = \sum_{\alpha,\beta}
\Delta {R}_{\alpha} \left( \sum_i  g_{i,\alpha}  g_{i,\beta} \right)  \Delta {R}_{\beta}
\end{equation}

where $\bm {F^0}$ is the fingerprint of the reference configuration, $\bm {F(R)}$ is the fingerprint of the displaced configuration and $g_{i,\alpha}$ is the gradient of the $i$-th component of the fingerprint vector with respect to 
the Cartesian components $\alpha$ of the position vector $\bm{R}$. 
\begin{equation}
    g_{i,\alpha} = \left.  \frac{ \partial F_i}{\partial R_{\alpha} } \right|_{\bm{R} = \bm{R}_0} \
\end{equation}
Hence $\alpha$ and $\beta$ run from 1 up to three times the number of atoms. Since we use four orbitals per atom (one s and 3 p type orbitals), $i$ runs from 1 to four times the number of atoms $N$.

The symmetric $3N \times 3N$  matrix $S_{\alpha,\beta}=\sum_i g_{i,\alpha}g_{i,\beta}$ is the sensitivity matrix. It tells us how strongly the fingerprint varies when the atoms are moved. $N$ is the number of neighbours of the reference atom (including itself) for which the sensitivity matrix is calculated. Since $\bm S$ is symmetric its eigenvalues are real and its eigenvectors form a complete basis set of the configurational space, \textit{i.e.}:
\begin{equation}
    \bm S=\sum_{i=1}^{3N} \lambda_i |\bm v_i><\bm v_i|
\end{equation}
Since we can write any arbitrary displacement as $\bm {\Delta R}=\sum_i c_i |\bm{v_i}>$ with $c_i=<\bm{v_i}|\bm {\Delta R}>$, the square of fingerprint distance can be written in terms of the eigenvalues contribution as:
\begin{equation}
    \left(\bm{F(R)} - \bm{F^0}\right)^2=\sum_i \lambda_i c_i^2
\end{equation}
This means that the eigenvectors of the sensitivity matrix are displacement modes and the eigenvalues show how much the fingerprint changes under movements along these modes. 
The sensitivity matrix has 6 zero eigenvalues. The associated eigenvectors describe 3 uniform translations and 3 rotations of the atoms. 
In the following we will only consider displacements that do not contain any translations or rotations.
For a unit displacement along the $i$-th eigenvector of the sensitivity matrix, the fingerprint distance $\Delta F$ is $\sqrt{\lambda_i}$. 
So if the sensitivity matrix has more than 6 zero eigenvalues there will be infinitesimal displacements that leave the fingerprint invariant. If more than 6 zero eigenvalues exist not only in a point but on a manifold, one can follow these zero modes by just moving by small amounts in the direction of 
the corresponding eigenvectors and obtain in this way finite displacements that leave the fingerprint invariant.  It is actually not necessary that the eigenvalue is exactly zero. If there exists a manifold on which one eigenvalue is 
smaller by several orders of magnitude than the other eigenvalues there will be finite movements 
that leave the fingerprint nearly constant compared to other movements. Fingerprints which are constant or numerically quasi-constant are clearly problematic. It means that such a fingerprint can not distinguish any more 
structural differences that give rise to different physical and chemical properties.

In the tests of all systems presented in this work we used $n_{max}=l_{max}=16$ as well as a Gaussian width $ \sigma $ of 0.3 \AA\ for SOAP. To check that the results are not improved by smaller values of $ \sigma$ we redo our  
calculations for $ \sigma = 0.1$ for one system. For ACSF we used 10 radial and 48 angular symmetry functions with standard parameters as in \cite{parsaeifard2020assessment}. A cutoff of 4.8 \AA\ is used for ACSF. 
For SOAP the cutoff is given by $\sigma n_{max}$. Our observations show that the appearance  of the manifold and its effect on machine learning do not change significantly for other reasonable choices of the parameters. We have scaled the sensitivity matrix for all fingerprints such that 
the largest eigenvalue is one.

\begin{figure*}[!tp]
    \centering
    \includegraphics[width=0.75\textwidth]{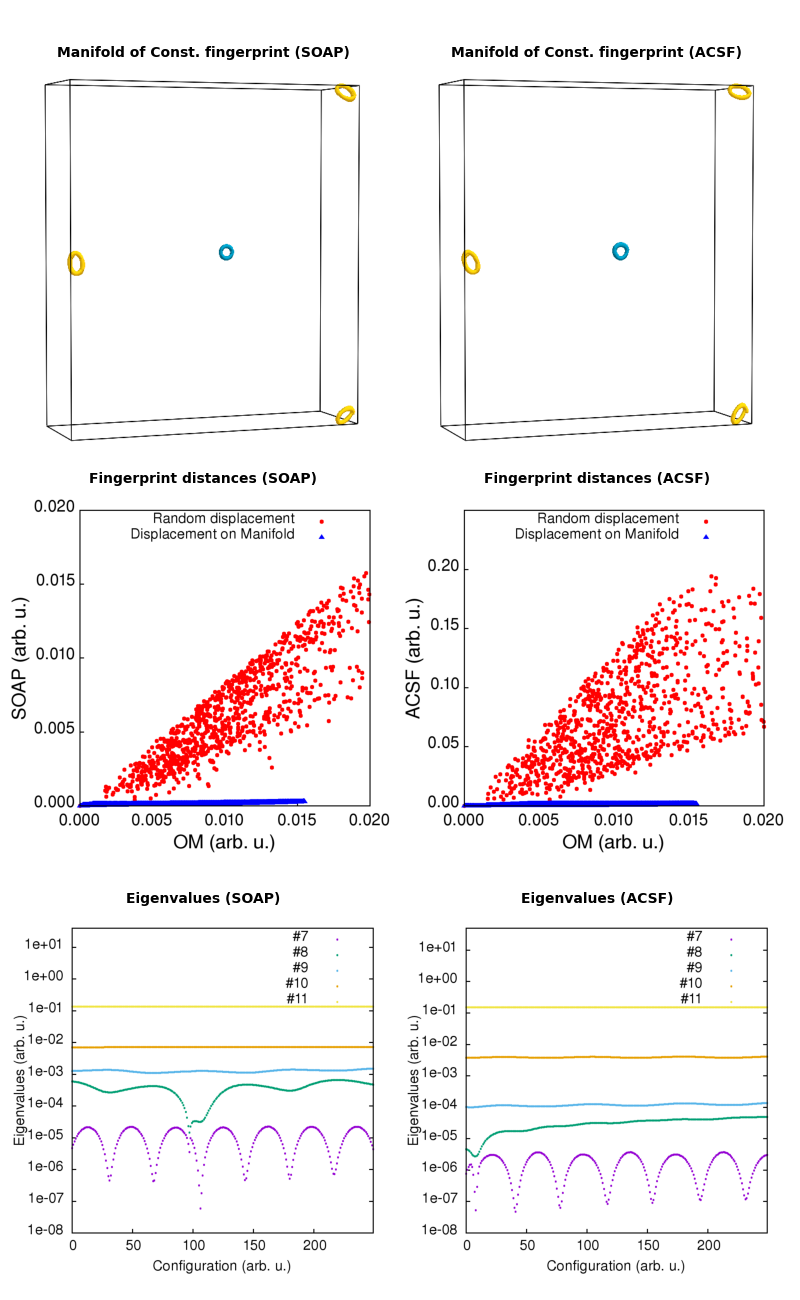}
    \caption{The top row shows the translations and rotation free manifold of quasi-constant fingerprint of the central N atom in NH$_3$ for SOAP (left column) and ACSF (right column). 
    The diameter of the largest rings in both SOAP and ACSF is about 0.1 \AA. The second row shows the fingerprint distances of the N atom for configurations along a trajectory on this manifold. The fingerprint distances obtained with OM are compared with the SOAP/ACSF distances. The red circles indicate the fingerprint distances of the N-atom in randomly displaced configurations (by a maximum of 0.02 \AA) and the blue triangles distances on the manifold. The last row shows the eigenvalues of the sensitivity matrix along the manifold. Eigenvalues 1 to 6, corresponding to uniform translations and rotations, are zero up to the machine precision level and are not shown. 
    The largest eigenvalue, which is always scaled to 1, is not shown either.}
    \label{fig:nh3}
\end{figure*}

\begin{figure*}[!tp]
    \centering
    \includegraphics[width=0.75\textwidth]{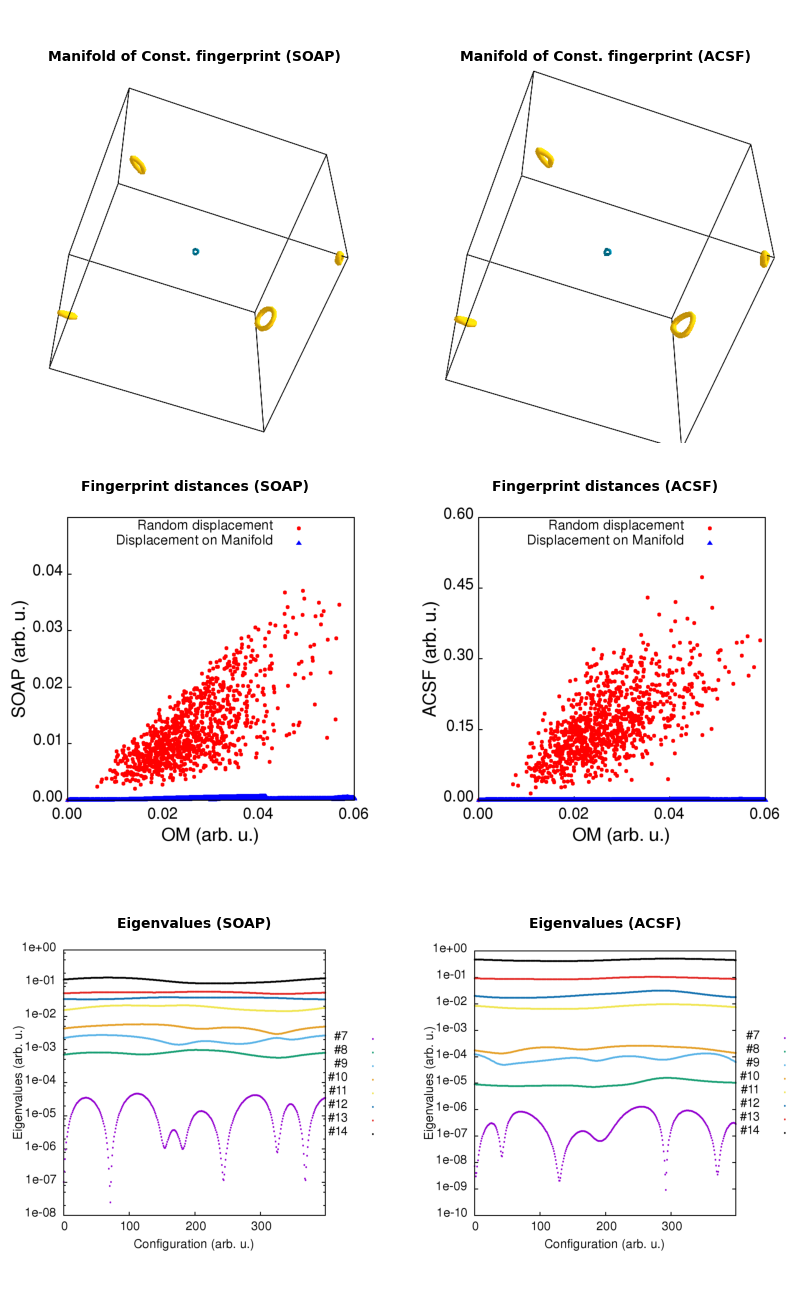}
    \caption{Same as Fig. \ref{fig:nh3} but for C atom in CH$_4$. The diameter of the largest rings is about 0.15 \AA.
    }
    \label{fig:ch4}
\end{figure*}

\section{results} \label{result}
\subsection{Manifolds of Quasi-Constant Fingerprint}
We apply now the method explained in the previous section to three small molecules, namely 
H$_2$O, NH$_3$, and CH$_4$ to investigate the manifold of constant fingerprint. 
The names of these molecules are in this context just placeholders for the scenarios where a central atom has environments containing 2, 3 or 4 neighboring atoms.
In the case of H$_2$O the reference atom O has only two neighbours, \textit{i.e.} any fingerprint with three-body terms is sufficient to fully characterize its local environment. As expected, no manifold of constant fingerprint can be found with SOAP and ACSF for the O-atom in H$_2$O. 

For the NH$_3$ and CH$_4$ where higher order many-body term come into play, the situation is however different and  constant fingerprint manifolds exists.
In Fig. \ref{fig:nh3} we show the SOAP constant fingerprint trajectory in the upper left panel for NH$_3$. The rings shown in the figure are the superpositions of many frames along such a trajectory that leaves the fingerprint invariant. In each frame the atoms are represented by small spheres such that the superposition of all the atomic positions describes 
the movement. 
As can be seen from the bottom left panel the smallest eigenvalue of the sensitivity matrix is in this case about 5 to 6 orders of magnitude smaller than the largest one. 
The blue triangles in the middle panels of the Fig. \ref{fig:nh3} represent the fingerprint distance between the N atom in the configurations along the trajectory on the manifold and the N atom in the initial configuration on the manifold. 
Even though 
the configurations undergo finite movements on these rings, the SOAP fingerprint distance is less than $5\times 10^{-4}$ among all the configurations on the ring. This is very small compared to random movements of comparable amplitude which 
lead to fingerprint distances of about $10^{-2}$ as shown by the red dots in Fig.~\ref{fig:nh3}. We could generate such quasi-constant manifolds for any 
initial structure that is not too close to a high symmetry structure. So this means that there is not only one 
such manifold but essentially an infinite number of them.

The right column of Fig. \ref{fig:nh3} shows the results for the ACSF. 
The manifold is also ring shaped and its diameter is very similar to the case of SOAP. 
The ratio between the fingerprint distances for equal amplitude movement on and off the manifold are also very similar. For the trajectory shown in  the top right panel of Fig.~\ref{fig:nh3} the smallest eigenvalue was 6 to 7 orders of magnitude smaller than the largest one and the fingerprint 
changed by less than $2\times 10^{-3}$ along the trajectory 
whereas for random displacements with similar amplitudes the fingerprint distance changed by up to $0.2$.



The same kind of manifolds exist for the carbon atom in CH$_4$.
Fig.~\ref{fig:ch4} shows the manifolds of constant fingerprint for the central C-atom in a CH$_4$ molecule with both SOAP and ACSF. 
The diameter of the largest ring is about 0.15 \AA\ for both SOAP and ACSF. 
As can be seen from the bottom panels in Fig~\ref{fig:ch4} the ratio of the largest to the smallest eigenvalue is about $10^5$ for SOAP and 
$10^7$ for ACSF and the fingerprints change by less than $2\times 10^{-3}$ for ACSF and less than $5\times 10^{-4}$ for SOAP.

 We couldn't find such a manifold for any of the studied molecules and configurations with the OM fingerprint, \textit{i.e.} the OM fingerprint always recognizes structural differences. For this reason we use the OM fingerprint in Fig.~\ref{fig:nh3} and Fig.~\ref{fig:ch4} to detect differences between the atomic environments on the manifold of the SOAP and ACSF fingerprints.

\begin{figure*}[tp!]
    \centering
    \begin{subfigure}[b]{\columnwidth}
         \centering
         \includegraphics[width=\textwidth]{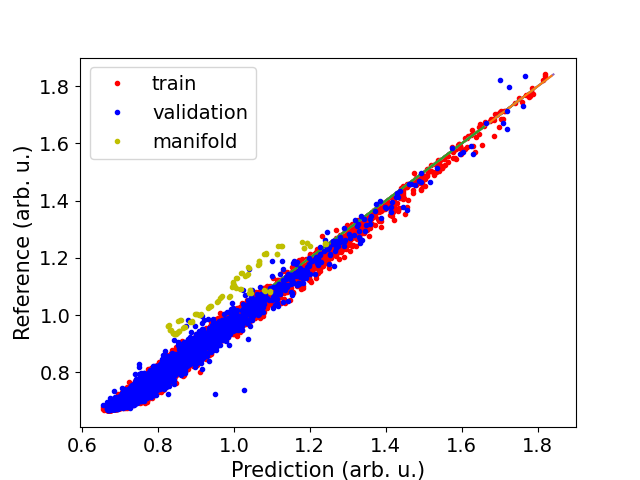}
         \caption{}
    \end{subfigure}
    \begin{subfigure}[b]{\columnwidth}
         \centering
         \includegraphics[width=\textwidth]{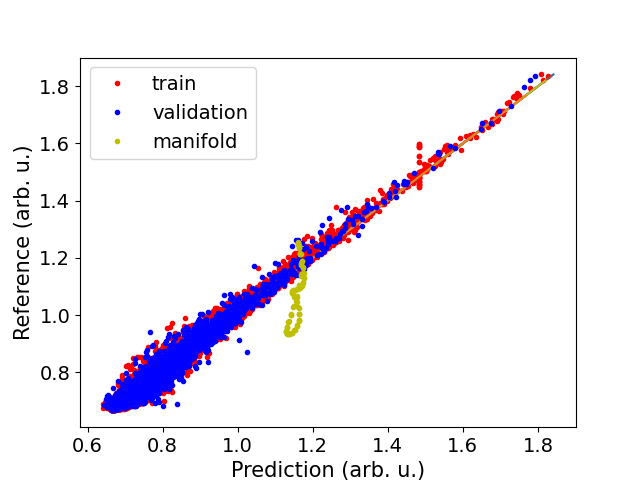}
         \caption{}
    \end{subfigure}
    \begin{subfigure}[b]{\columnwidth}
         \centering
         \includegraphics[width=\textwidth]{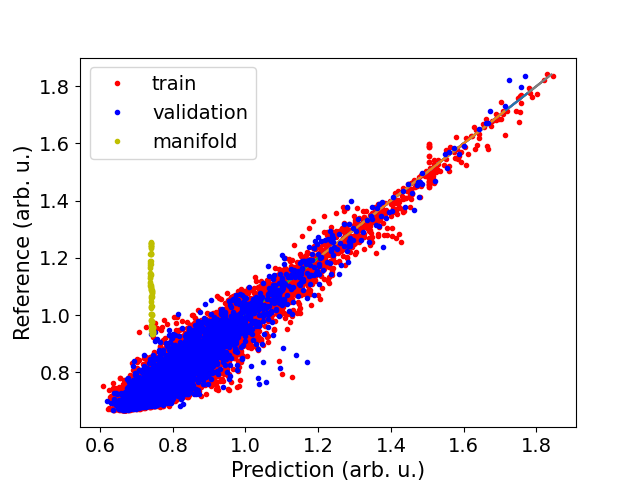}
         \caption{}
    \end{subfigure}
    \caption{ Machine learning results of four-body energies for \textbf{a}) OM, \textbf{b}) SOAP, and \textbf{c}) ACSF. The SOAP and ACSF fingerprints have a very poor extrapolation for the four-body energies of the structures on the manifold. }
    \label{fig:ML}
\end{figure*}

\begin{figure*}[tp!]
    \centering
    \begin{subfigure}[b]{\columnwidth}
         \centering
         \includegraphics[width=\textwidth]{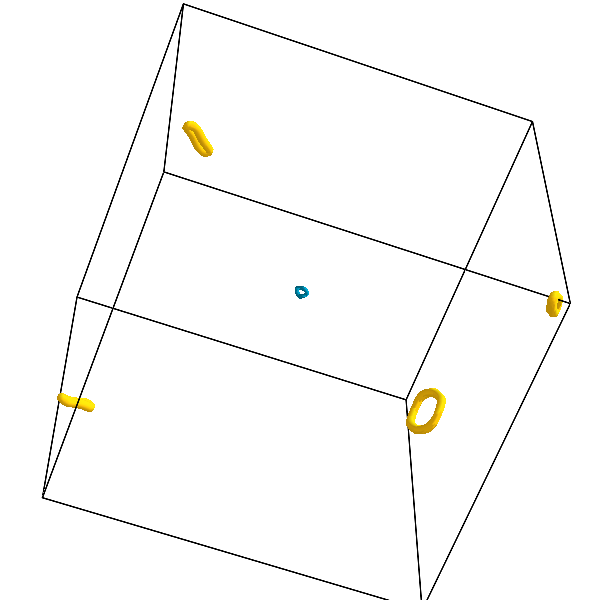}
         \caption{}
    \end{subfigure}
    \begin{subfigure}[b]{\columnwidth}
         \centering
         \includegraphics[width=\textwidth]{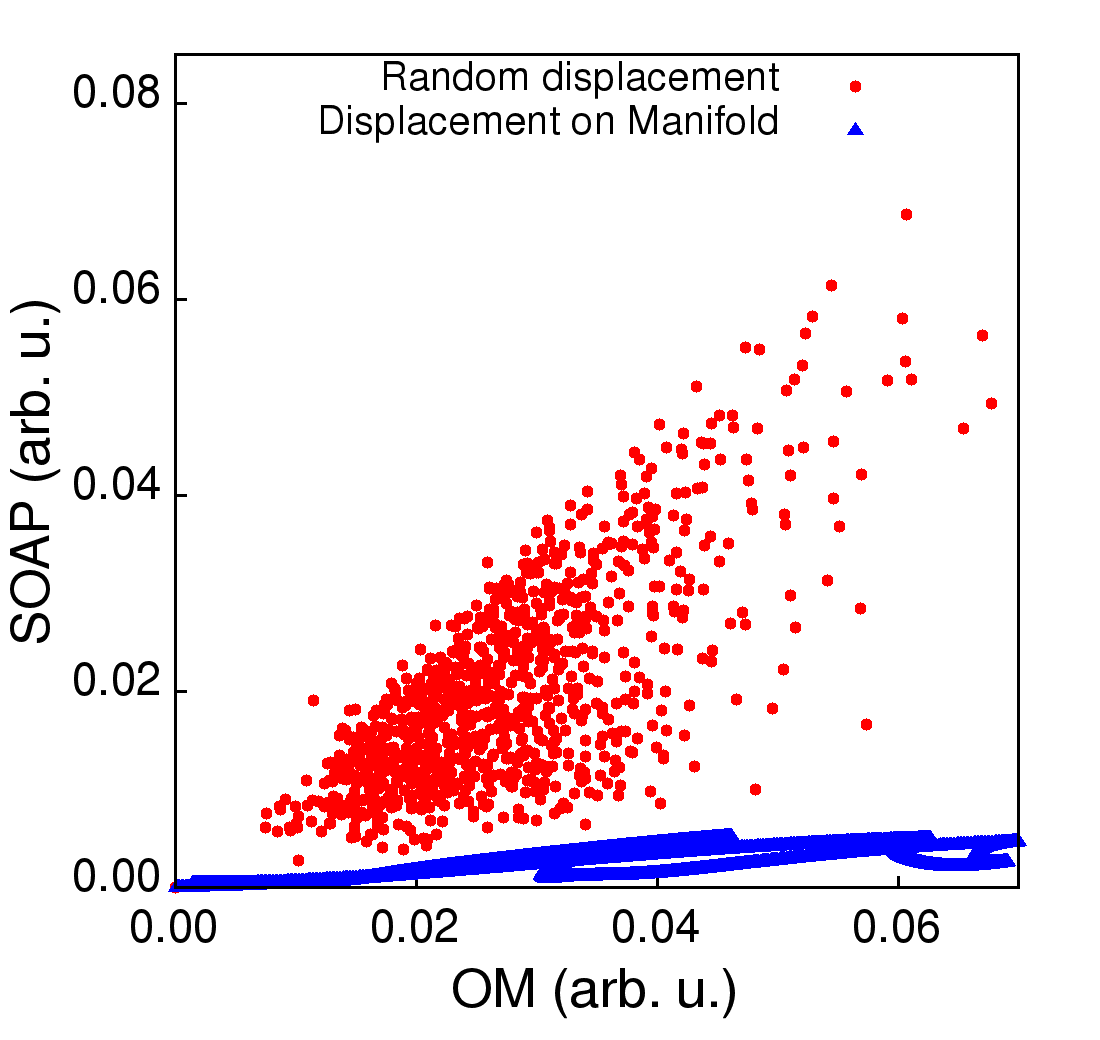}
         \caption{}
    \end{subfigure}
    \begin{subfigure}[b]{\columnwidth}
         \centering
         \includegraphics[width=\textwidth]{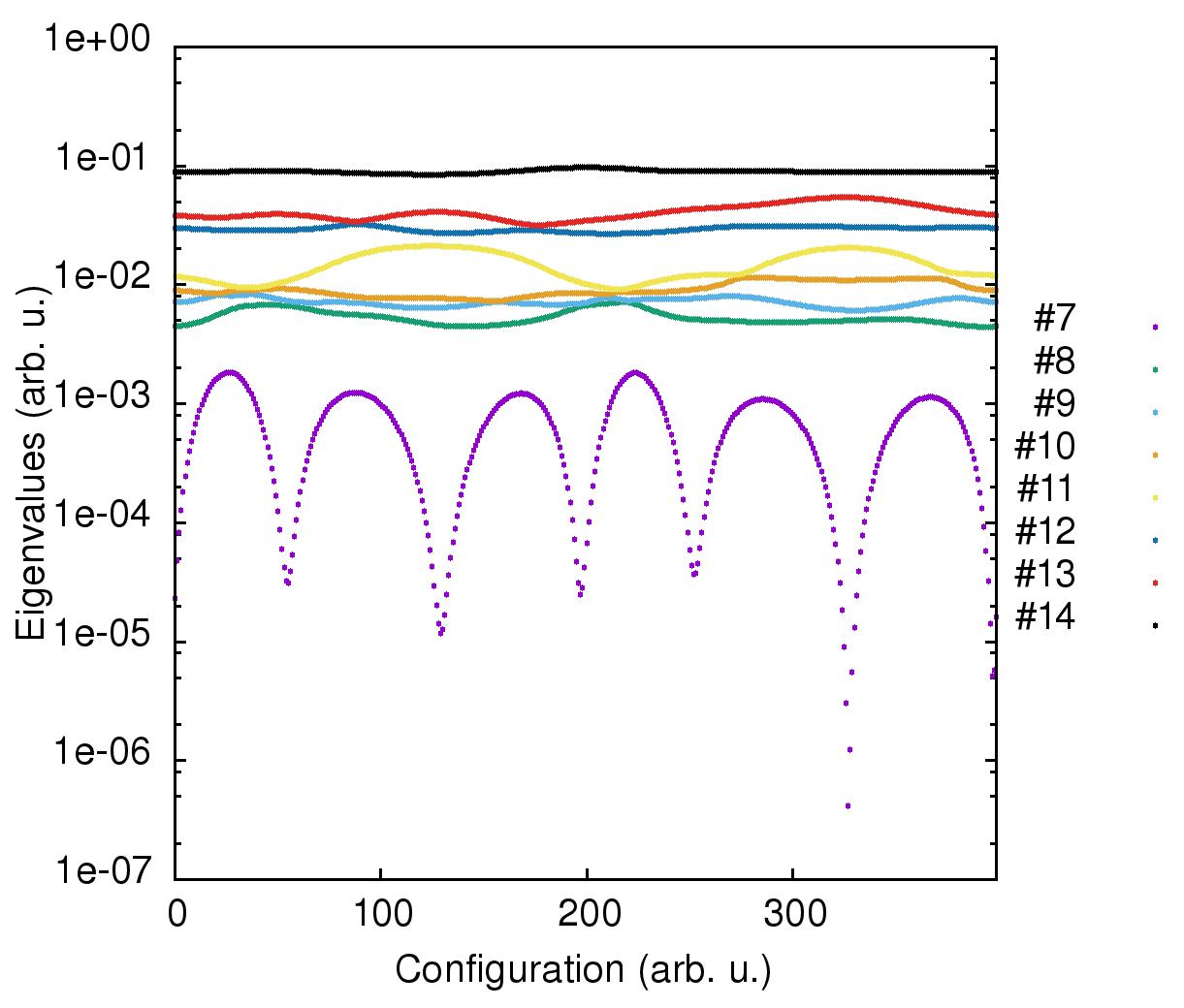}
         \caption{}
    \end{subfigure}
    \begin{subfigure}[b]{\columnwidth}
         \centering
         \includegraphics[width=1.1\textwidth]{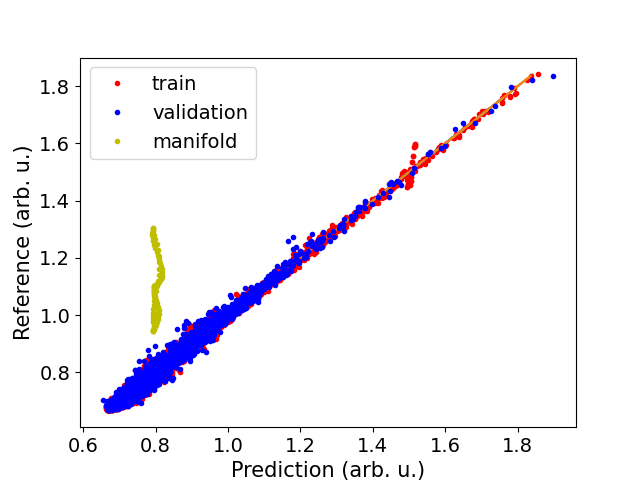}
         \caption{}
    \end{subfigure}
    \caption{ \textbf{a}) The manifold of constant fingerprint \textbf{b}) the fingerprint distances, \textbf{c}) The eigenvalues of the sensitivity matrix along the manifold, and \textbf{d}) the machine learning results of four-body energies for the SOAP fingerprint with $\sigma=0.1$. The result shows that lowering $\sigma$ does not eliminate the manifolds of constant fingerprint for SOAP.
    Even though some erratic variation of the fingerprint can now be seen by eye in panel \textbf{b}), this variation is still too small 
    to suppress the problems in machine learning (panel \textbf{d}).}
    \label{fig:ML-soap0.1}
\end{figure*}

\subsection{Consequences for Machine Learning}

\begin{figure}[h!]
    \centering
    \includegraphics[width=\columnwidth]{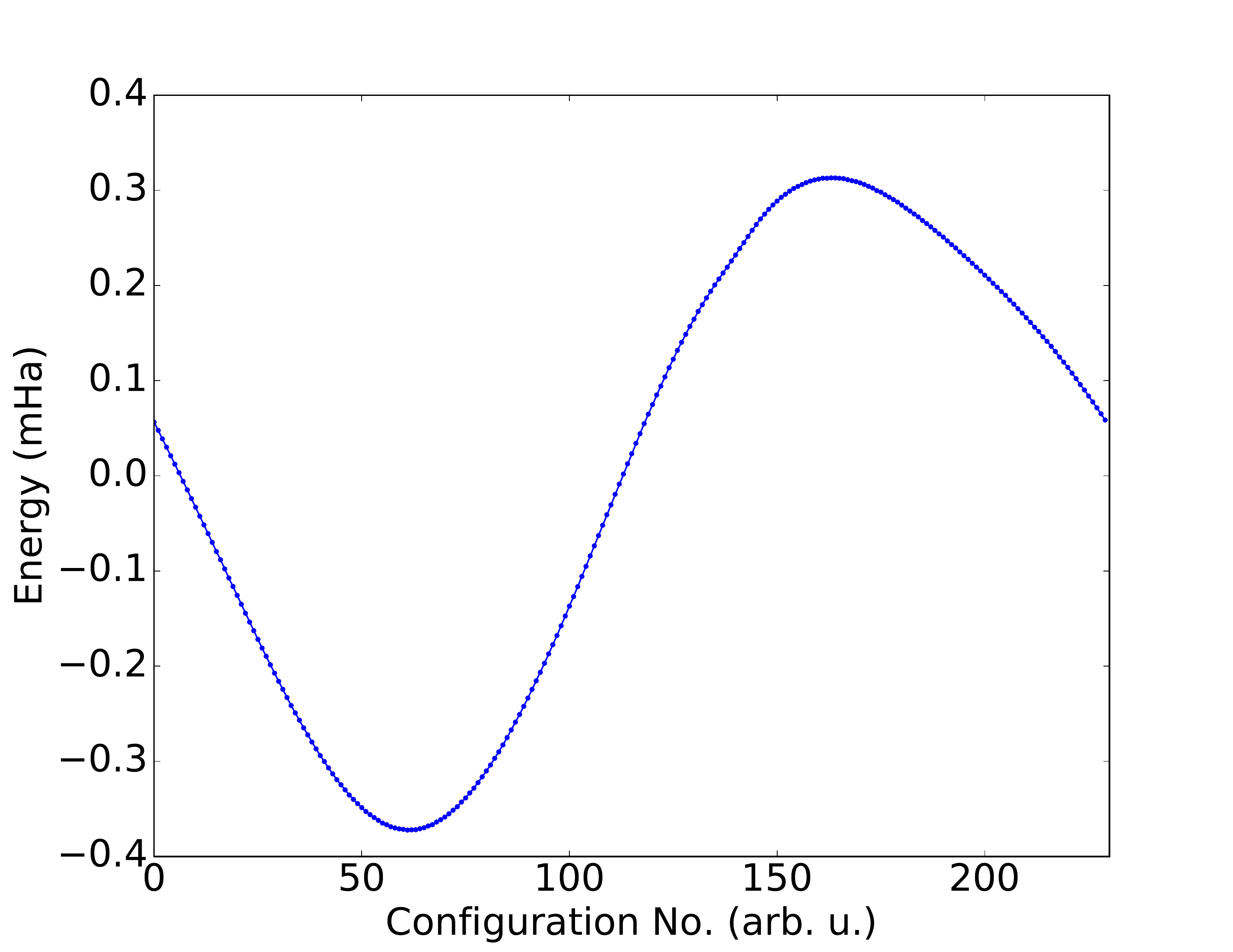}
    \caption{Total DFT energy of  CH$_4$ structures evenly distributed on the manifold of quasi-constant fingerprint of Fig.\ref{fig:ch4}. The energy variation is of the order of a mHa.  }
    \label{fig:en-along-manifold}
\end{figure}

The bond lengths between the central atom and the surrounding atoms change
hardly for the movements on the manifolds. The angles between the central atom and pairs of surrounding atoms also change very little.
Since the two and three body terms are responsible for the largest variation of the energy, the DFT energy varies only by about a mHa 
on these manifolds (Fig.~\ref{fig:en-along-manifold}). 
But even though this energy change is small, tortional energies of this 
order are responsible for the the folding of proteines and other 
biomolecules. Therefore such four-body interactions are considered an essential part in all classical force fields. 
An energy variation of a mHa is also about one order of magnitude larger than the target accuracy of modern machine learning schemes which is one meV/atom~\cite{behler_machine_2021} i.e. about 0.04 mHa/atom.

The deficiencies of the SOAP and ACSF fingerprints in the context of machine learning can therefore best be detected if one considers only four-body terms. For this reason we consider a restricted energy that consists only of the standard torsion terms found in most force fields.
It consists of the sum of all cosine squared of the torsion angles with respect to the central atom.
As expected and shown in Fig. \ref{fig:ML}, a machine learning scheme based on the SOAP and ACSF fingerprints for the central atom is not able to learn this four-body energy for structures on a quasi-constant manifold. Actually even if the configurations of the manifold 
are included in the training set they can not be machine learned. 
Since there is a huge number of manifolds, it would in practice however anyway not be possible to include all manifolds in the training set.  If, in contrast, the OM fingerprint is used, the energy can be learned without the need to include points on manifolds. 

It was recently shown that the condition number, i.e the  ratio between the largest and smallest eigenvalue of the mapping from cartesian 
coordinates to a SOAP fingerprint vector is improved for values of 
$\sigma$ that are smaller than the values used in previous applications~\cite{mafia}. 
This implies that the smallest eigenvalues of the sensitivity matrix 
for configurations on the manifold should be larger for smaller values of $\sigma$. For this reason we have repeated our calculations of $CH_4$ with 
a smaller value of $\sigma = 0.1$. We can indeed observe that the smallest non-zero eigenvalue is now larger, but the manifold is hardly changed (see Fig.~\ref{fig:ML-soap0.1}).
In addition, the variation of the fingerprint on the manifold is still so small 
that machine learning of the manifolds fails (see Fig~\ref{fig:ML-soap0.1}). 

 The machine learning results shown in this paper were obtained by the Pytorch~\cite{paszke2017automatic, NEURIPS2019_9015} software using a feed-forward neural network with three hidden layers of 64, 128, and 64 nodes. 
 The activation function for the output layer was linear and for all other layers 
 the tanh was used. A batch normalization on the hidden layers was also performed. We used the Adam optimizer \cite{kingma2014adam} for the optimization of the network. The training was continued until the loss function could not be improved any further.

\section{Conclusion}
In conclusion, by following the eigenvectors of the sensitivity matrix corresponding to small eigenvalues we can find manifolds of quasi-constant fingerprint for SOAP and ACSF. The fact that configurations that are different and have therefore different torsional energies, are considered to be quasi-identical by the fingerprint prevents the machine learning of these torsional energies. 
The four-body interactions that are explicitly included in all traditional force fields
can therefore only be reproduced with limited accuracy in a machine learning scheme that uses these fingerprints.
Contrary to a widespread belief the SOAP fingerprint is not better than ACSF in resolving four-body terms. 
The many-body nature of the OM fingerprint makes it an attractive fingerprint for resolving with high fidelity structural differences as well as for various applications~\cite{AliML,joost}.

\section{Acknowledgements}
This research was partly performed within the NCCR MARVEL, funded by the Swiss National Science Foundation. The calculations were performed on the computational resources of the Swiss National Supercomputer (CSCS) under project s963 and on the Scicore computing center of the University of Basel.

\section{conflict of interest}
The authors have no conflicts to disclose.

\section{DATA AVAILABILITY}
The data that support the findings will be made available after acceptance of this MS on github.



\bibliography{main}

\end{document}